\documentclass[sigconf]{acmart}

\setcopyright{none}
\renewcommand\footnotetextcopyrightpermission[1]{}
\AtBeginDocument{%
  \providecommand\BibTeX{{%
    Bib\TeX}}}
    
\usepackage{endnotes}
\usepackage{color}
\usepackage{graphicx}
\usepackage{wrapfig}
\usepackage{multirow}
\usepackage{listings}
\usepackage{url}
\usepackage{fixltx2e}
\usepackage{balance}
\usepackage{listings}
\usepackage{verbatim}
\usepackage{subfigure}
\usepackage{stfloats}
\usepackage{amsmath}
\usepackage{tikz}

\def\BibTeX{{\rm B\kern-.05em{\sc i\kern-.025em b}\kern-.08em
    T\kern-.1667em\lower.7ex\hbox{E}\kern-.125emX}}

\definecolor{mGreen}{rgb}{0,0.6,0}
\definecolor{mGray}{rgb}{0.5,0.5,0.5}
\definecolor{mPurple}{rgb}{0.58,0,0.82}
\definecolor{backgroundColour}{rgb}{0.95,0.95,0.92}

\lstdefinestyle{CStyle}{
    backgroundcolor=\color{backgroundColour},   
    commentstyle=\color{mGreen},
    keywordstyle=\color{magenta},
    numberstyle=\tiny\color{mGray},
    stringstyle=\color{mPurple},
    basicstyle=\ttfamily\footnotesize,
    breakatwhitespace=false,         
    breaklines=true,                 
    captionpos=b,                    
    keepspaces=true,                 
    numbers=left,                    
    numbersep=5pt,                  
    showspaces=false,                
    showstringspaces=false,
    showtabs=false,                  
    tabsize=2,
    language=C
}

\newif\ifdraft
\drafttrue
\draftfalse                                                                              
\ifdraft
\newcommand{\zhao}[1]{{\textcolor{cyan}    { ***Zhao:      #1 }}}
\newcommand{\katznote}[1]{{\textcolor{magenta}    { ***Dan:      #1 }}}
\newcommand{\outline}[1]{{\textcolor{blue}    { ***Outline:      #1 }}}
\newcommand{\lei}[1]{{\textcolor{red}    { ***Lei:      #1 }}}
\newcommand{\revision}[1]{ {\textcolor{red}    {\bf #1 }}}
\definecolor{darkgreen}{rgb}{0,0.7,0}
\newcommand{\groppnote}[1]{{\textcolor{darkgreen}    { ***Bill:      #1 }}}
\else
\newcommand{\zhao}[1]{}
\newcommand{\katznote}[1]{}
\newcommand{\groppnote}[1]{}
\newcommand{\outline}[1]{}
\newcommand{\lei}[1]{}
\newcommand{\revision}[1]{}
\fi

\newenvironment{shortlist}{
        \vspace*{-0.5em}
  \begin{itemize}
  \setlength{\itemsep}{-0.1em}
}{
  \end{itemize}
        \vspace*{-0.5em}
}
\newcommand{\up}{\vspace*{-0.5em}}

\newcommand{\name}{ThemisIO}

\begin{document}

\title{Fine-grained Policy-driven I/O Sharing for Burst Buffers}

\author{Ed Karrels$^*$}
\affiliation{
    \institution{University of Illinois Urbana-Champaign} 
    \country{}
}
\email{edk@illinois.edu}

\author{Lei Huang$^*$}
\affiliation{
    \institution{Texas Advanced Computing Center} 
    \country{}
}
\email{huang@tacc.utexas.edu}

\author{Yuhong Kan}
\affiliation{
    \institution{The University of Texas at Austin} 
    \country{}
}
\email{kan_yuhong@utexas.edu}

\author{Ishank Arora}
\affiliation{
    \institution{The University of Texas at Austin} 
    \country{}
}
\email{ishankarora1100@utexas.edu}

\author{Yinzhi Wang}
\affiliation{
    \institution{Texas Advanced Computing Center} 
    \country{}
}
\email{iwang@tacc.utexas.edu}

\author{Daniel S.\ Katz}
\affiliation{
    \institution{University of Illinois Urbana-Champaign} 
    \country{}
}
\email{d.katz@ieee.org}

\author{William D.\ Gropp}
\affiliation{
    \institution{University of Illinois Urbana-Champaign} 
    \country{}
}
\email{wgropp@illinois.edu}

\author{Zhao Zhang}
\affiliation{
    \institution{Texas Advanced Computing Center} 
    \country{}
}
\email{zzhang@tacc.utexas.edu}

\begin{abstract}
A burst buffer is a common method to bridge the performance gap between the I/O needs of modern supercomputing applications and the performance of the shared file system on large-scale supercomputers. 
However, existing I/O sharing methods require resource isolation, offline profiling, or repeated execution that significantly limit the utilization and applicability of these systems.
Here we present \name{}, a policy-driven I/O sharing framework for a remote-shared burst buffer: a dedicated group of I/O nodes, each with a local storage device.
\name{} preserves high utilization by implementing opportunity fairness so that it can reallocate unused I/O resources to other applications.
\name{} accurately and efficiently allocates I/O cycles among applications, purely based on real-time I/O behavior without requiring user-supplied information or offline-profiled application characteristics.
\name{} supports a variety of fair sharing policies, such as user-fair, size-fair, as well as composite policies, e.g., group-then-user-fair.
All these features are enabled by its statistical token design. \name{} can alter the execution order of incoming I/O requests based on assigned tokens to precisely balance I/O cycles between applications via time slicing, thereby enforcing processing isolation.
Experiments using I/O benchmarks show that \name{} sustains 13.5--13.7\% higher I/O throughput and 19.5--40.4\% lower performance variation than existing algorithms.
For real applications, \name{} significantly reduces the slowdown by 59.1–99.8\% caused by I/O interference.
\end{abstract}

\maketitle
\pagestyle{plain}
\def\thefootnote{*}\footnotetext{Equal contribution to this work}\def\thefootnote{\arabic{footnote}}

\section{Introduction}
High-performance computing (HPC) architects have deployed burst buffers on supercomputers to bridge the I/O gap between compute and storage by absorbing bursty I/O. 
However, when burst buffers are shared among applications, researchers have observed I/O interference~\cite{herbein2016scalable, kougkas2016leveraging, thapaliya2016managing, mubarak2017quantifying}, where one application is slowed down by the I/O of other applications.

\begin{figure}[t]
    \includegraphics[width=\columnwidth]{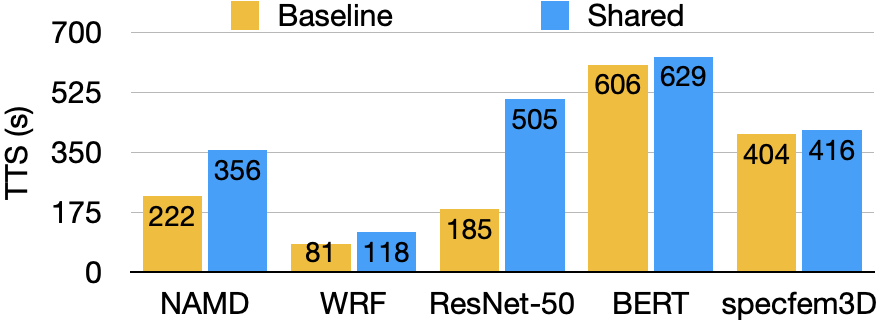}
    \caption{Time-to-solution comparison of five applications using a remote burst buffer. Baseline = measurements with exclusive access to the burst buffer. Shared = measured with a background I/O benchmark job running. }
    \label{fig:interference}
\end{figure}

To illustrate this, we run a set of five applications in a controlled environment with an in-house remote-shared burst buffer using two nvdimm nodes on the Frontera supercomputer. Each node is equipped with a 6.2~TB Intel Optane persistent memory. 
We first measure the baseline performance of each application making exclusive use of the burst buffer, then we measure each application's runtime with a background I/O benchmark job, which is 3--173\% longer than the baseline, as shown in Figure~\ref{fig:interference}.

The root cause of I/O interference is that today’s production systems generally process I/O requests in a first-in-first-out (FIFO) manner, which means that highly concurrent and bursty I/O traffic from one application can saturate the I/O system's queue, then block the I/O of another application for a period of time. 
The amount of the slowdown depends on the state of the queue when I/O requests enter; specifically, how many I/O requests from one application are ahead of the new requests from the other application.

Computer scientists have proposed various ideas to enable efficient I/O sharing~\cite{qian2017configurable, patel2020gift, ji2019automatic, herbein2016scalable, kougkas2016leveraging, liang2019cars} .
However, these solutions require prior knowledge of the application's I/O behavior to enable fair-sharing of I/O resource.
E.g., Gift~\cite{patel2020gift} assumes applications are repeatedly run, it uses a throttle-and-reward system to maximize I/O bandwidth by relaxing the fairness window.
This throttled execution in Gift may introduce a unbounded delay.
TBF~\cite{qian2017configurable} and CARS~\cite{liang2019cars} require users to supply the I/O throughput of applications, then they statically allocate sufficient I/O resource to meet the requirement.
Because I/O workload can vary during an application's execution, a static allocation can lead to wasted I/O resources.
These systems cannot dynamically adjust I/O resource allocations based on the real-time I/O requirement for new applications, or even existing applications with new configurations, e.g., running at a larger scale. 

To address the I/O interference issue in burst buffer systems and to enable {\bf efficient} and {\bf dynamic} I/O resource sharing with {\bf flexible policies}, we have designed, built, and tested \name{}. 
To ensure \name{} is always operating with {\bf maximal I/O throughput}, 
it implements opportunity fairness, meaning that fairness is only enforced when the overall I/O requests received by the I/O system exceeds its capacity. 
While using \name{} when the I/O system is partially loaded, applications will get the same amount of I/O resources as they would when running without \name{} (see \S\ref{sec:exp:policy:prim}).
Thus, the delay of an applications due to enforced fairness is bounded, as \name{} guarantees the assigned I/O resources for each application are no less than its fair share as specified in a policy (see \S\ref{sec:exp:apps}). 
We have designed a statistical, token-based strategy for time-sliced sharing, which offers every I/O-active application an opportunity for fair-sharing, unlike previous work that uses a mandatory I/O bandwidth assignment. 
To enable \name{} to {\bf dynamically} adjust  I/O resources to applications based on their real-time workloads, we embed job-related information, such as job id, user id, and job size, in the I/O request.
\name{} can allocate or reclaim tokens based on the number of I/O requests from each user, each job, or each group (see \S\ref{sec:exp:compare}).
This token-based design also makes \name{} {\bf flexible in sharing polices}.
For example, \name{} can assign the same number of tokens to each user to achieve primitive sharing policies, such as user-fair (i.e., I/O cycles are evenly split across users.) 
It can also assign an appropriate number of tokens to enable composite policies such as user-then-job-fair, i.e., the I/O cycles are evenly split across users, then for each user, the allocated cycles are further evenly split among their jobs (see \S\ref{sec:exp:policy:com}).

To make \name{} compatible with existing supercomputing applications, we implement a POSIX-compliant interface, so applications can take advantage of \name{} as a tradition file system;
users do not have to make changes to their code to use \name{}.

We have integrated \name{} with an in-house burst buffer system and
used I/O benchmarks and five real-world applications to examine the efficacy of the \name{} design. 
Our results show that \name{} is efficient, as it achieves the hardware I/O throughput limit, which is $\sim$22~GB/sec per I/O server combining read and write. 
%The generic design of \name{} easily enables primitive sharing policies such as job-fair (i.e., I/O cycles are evenly split across jobs) and user-fair as well as composite policies of user-then-job-fair.
%\name{} can precisely and efficiently allocate I/O resources with the specified sharing policy. 
Global fairness can be preserved with a controlled delay, which is negligible compared to the time-to-solution of jobs (see \S\ref{sec:exp:global}).
The sharing enabled by \name{} is more efficient and stable than that of existing frameworks GIFT and TBF. 
With \name{}, the sustained I/O throughput of benchmarks is 13.5--13.7\% higher than when using the GIFT and TBF algorithms,
and the I/O throughput variation in I/O sharing is 19.5--40.4\% lower with \name{}.

For real applications, running a 64-node NAMD job with a background I/O benchmark job is 37.7\% faster with \name{} size-fair policy than that of FIFO.
Compared to the baseline with exclusive I/O resource access, the size-fair policy is 0.1\% slower while the FIFO policy is 60.3\% slower, which is a $\sim$600x reduction in the performance slowdown due to I/O interference.
Across the applications in \S\ref{sec:exp:apps}, \name{} size-fair reduces the slowdown by 59.1--99.8\% compared to FIFO.

The contributions of this paper are:
\begin{shortlist}
  \item{the statistical token-based time sharing design of \name{}, which can flexibly support various upstream sharing policies,}
  \item{the global fairness enforcement algorithm on multiple I/O servers, and} 
  \item{the design of primitive and composite I/O sharing policies.}
\end{shortlist}

The rest of the paper is structured as follows.
Section~\ref{sec:back} provides background on modern supercomputing applications and states the need and requirements of a policy-driven I/O sharing system.
The foundational statistical token is presented in \S\ref{sec:design-token}.
We then discuss the system design and implementation in \S\ref{sec:design} and  present experiments and their results in \S\ref{sec:experiments}.
Existing I/O sharing work is summarize and compared to \name{} in \S\ref{sec:related}.
Finally, we conclude and envision future work in \S\ref{sec:conc}.

\section{Background and Motivation}
\label{sec:back}
%Modern supercomputing applications exhibit intensive I/O, and researchers have explored various techniques to address the I/O challenge.
In this section, we first review some modern supercomputing applications and present a suite that we have selected to drive the design of \name{}, then discuss the motivation for I/O sharing.

\subsection{Background}
\label{sec:back:apps}
In addition to traditional numerical simulation applications, such as WRF (Weather Research and Forecasting)~\cite{skamarock2001prototypes} and NAMD (scalable molecular dynamics )~\cite{phillips2005scalable}, researchers now also use a big data approach (e.g., parallel scripting and MapReduce) and deep learning tools to conduct disciplinary research.
%Representative parallel scripting applications include the astronomical image mosaic application (Montage)~\cite{jacob2009montage} and the basic local alignment search tool (BLAST)~\cite{mathog2003parallel}. 
%One recent significant result is the groundbreaking LIGO gravitational wave discovery~\cite{abbott2016observation}.
Researchers also train neural networks for image classification, object detection, and scientific literature analysis. 
Some notable models ResNet~\cite{he2016deep}, Mask R-CNN~\cite{he2017mask}, and BERT~\cite{devlin2018bert}. 
Those models are being actively applied to research fields from astronomy to zoology~\cite{Charnock2016, kates2019predicting, casalino2020ai}.

These diverse applications, run daily on supercomputers, use I/O libraries such as MPI-IO~\cite{thakur1999implementing, thakur1997users}, HDF5~\cite{Folk1999hdf5}, and POSIX IO.
When running at large scale, some individual executions can cause file system saturation and unresponsiveness.
To absorb the bursty I/O workload, supercomputer architects have deployed burst buffer systems on supercomputers, such as the Cray DataWarp on NERSC's Cori and DDN Infinite Memory Engine (IME) on the OSC Pitzer and Owens clusters. 
The system we use here, Frontera, features 16 compute nodes that each have $\sim$5.4~TB of Intel Optane memory, configured as 2.1~TB memory extension and 3.3~TB local storage.

There are generally two types of burst buffers.
In a node-local burst buffer, a storage device such as an SSD is attached to each compute node, such as the NVDIMM nodes on Frontera.
\name{} is designed for the second type, remote-shared burst buffer, where storage devices are attached to a number of dedicated I/O nodes, such as  DataWarp on Cori.
In both cases, compute nodes and I/O nodes are connected via a high-speed interconnect, e.g., InfiniBand.
%In addition to traditional communication, NVMe over fabrics allows storage devices to communicate with each other without involving the host CPU~\cite{minturn2015nvm}.

\begin{comment}
To design \name{}, we use five representative applications (summarized in Table~\ref{tb:apps}) to compose a mixed workload that approximates actual supercomputer workloads. 

\begin{table}
\caption{Selected Application Summary \label{tb:apps}}
\begin{center}
    \begin{scriptsize}
    \begin{tabular}{ | c | c | c | c |}
    \hline
    Name & cientific Domain & I/O Lib & Data size  \\ \hline \hline
    WRF~\cite{skamarock2001prototypes} &weather forecast & POSIX, NetCDF & $O(10)$~GB \\ \hline
    NAMD~\cite{phillips2005scalable} & molecular dynamics & POSIX &$O(20)$~GB \\ \hline
    SPECFEM3D~\cite{bie2016towards} & seismology & POSIX & $O(1)$~GB \\ \hline
    %Montage & Big Data & astronomy & POSIX & $O(10)$~GB \\ \hline
    %COMPASS~\cite{abbon2007compass} & Big Data & high energy physics & $O(100)$~GB \\ \hline
    ResNet-50~\cite{he2016deep} & computer vision & POSIX & $O(100)$~GB \\ \hline
    %Mask R-CNN~\cite{he2017mask} & DL & computer vision & $O(10)$~TB \\ \hline
    BERT~\cite{devlin2018bert} & language modeling & POSIX, HDF5 & $O(100)$~TB \\ \hline
    \end{tabular}
    \end{scriptsize}
\end{center} 
\end{table}
\up\up
\end{comment}

\subsection{Motivation}
\label{sec:back:motivation}
Most of today's supercomputers provide processing isolation for computing resources by granting exclusive access to compute nodes. 
However, such isolation does not exist in I/O resources, which causes problems. 
In this section, we discuss these problems in detail and provision a multi-tenant I/O  framework to address I/O sharing challenges. 

\subsubsection{State of the Practice and Problems}
From our daily experience of operating HPC systems at multiple institutions, we have noticed cases where the I/O of a small job can dominate the systems I/O resources due to its high frequency and high volume.
%Such affects can originate from a small number of compute nodes and slow down all other jobs in the system.
The root causes of this phenomena are that 1) the I/O queue is packed with requests from the small job and the I/O system processes them in a FIFO way, and 2) the file system's throughput is insufficient. 
Depending on when other I/O requests arrive in the I/O queue, the small job can indefinitely block the I/O requests of all other jobs.
If we define fairness of I/O resources to be proportional to a job's size, i.e., the number of compute nodes, 
{\bf this unfair sharing of the I/O resource certainly does not reflect the priority of the jobs.}

In some other cases, the I/O workload of a job can be heavy in metadata access, which eventually saturate the metadata server. 
While this blocks other jobs from accessing metadata, the data servers, such as OSTs in Lustre, may be idle during this process.
Again, {\bf it is the FIFO processing of I/O requests that causes this huge resource waste}. 
Although using burst buffers can absorb the bursty workload to some extent, the problem still exists as long as the I/O requests are processed in a FIFO manner.

{\bf It is challenging, if not impossible, for today's I/O systems to provide a quality of service (QoS) guarantee}. 
For example, an I/O and shared file system that can provide 10M IOPS typically cannot guarantee 1M IOPS to each of ten jobs on the machine, regardless of the job's size.
One solution to this is to disassociate I/O control from actual processing.
With rich job metadata, control logic can alter the processing order of the I/O requests to enforce defined sharing policies and achieve fairness.

Those practices and problems motivate \name{}. 
Clearly, there is a critical need for an I/O sharing framework that can 1) isolate I/O request processing so one job does not block others, 2) assign idle I/O cycles to jobs with high I/O demand when possible to achieve high utilization, and 3) provide flexible sharing policies to enforce certain fairness, as discussed in detail in the next subsection.

\subsubsection{Definition of Fairness}
\label{sec:motivation-def}
There are many ways to define fairness. 
Allocating I/O resources in proportion to job size is one type of fairness, which we refer to as {\bf size-fair}.
Evenly splitting I/O resources among all active jobs is another type of fairness, and we refer to this as {\bf job-fair}.
Splitting I/O resources among all users regardless of the number of jobs each user is running is a third type of fairness, which we call {\bf user-fair}.

These are just three examples of primitive fairness definitions. 
One can argue that assigning more I/O resources to prioritized jobs is fair, for example, during the hurricane season where there is an urgent need for computing resources to forecast potential disasters;
we refer to this as {\bf priority-fair}.

Some cases may need composite sharing policies, for example, to first split I/O resources evenly among users, then to split a user's resources in proportion to their job sizes. We call this {\bf user-then-size-fair}.
There could also be {\bf group-then-size-fair} and {\bf group-then-user-fair} policies.

The \name{} design can support all of these sharing policies with a single parameter, so that system administrators can specify the sharing policy when starting \name{}. 
Internally, \name{} uses a token-based design to enable time-sliced sharing of I/O resources. 
It maps the above sharing polices and fairness definitions to token management, where
I/O requests are processed with corresponding job tokens, and tokens are recycled after requests are processed.
This means that sharing policies can be implemented by assigning a number of tokens to each job. 
The details of the token-based design is discussed in \S\ref{sec:design-token}.

\section{Sharing via Statistical Tokens}
\label{sec:design-token}
We designed \name{} to be a generic I/O sharing system, in the sense that it can flexibly support {\bf primitive sharing policies} as well as {\bf composite sharing policies}.
For primitive policies, e.g., {\bf job-fair}, {\bf user-fair}, {\bf size-fair}, and {\bf priority-fair}, we  assign an identical number of I/O cycles among jobs, users, or in proportion to the node count or the job priority within a time unit.
We refer the notion of job, user, and size as sharing entities. 

\begin{figure}[h]
\begin{center}
    \includegraphics[width=\columnwidth]{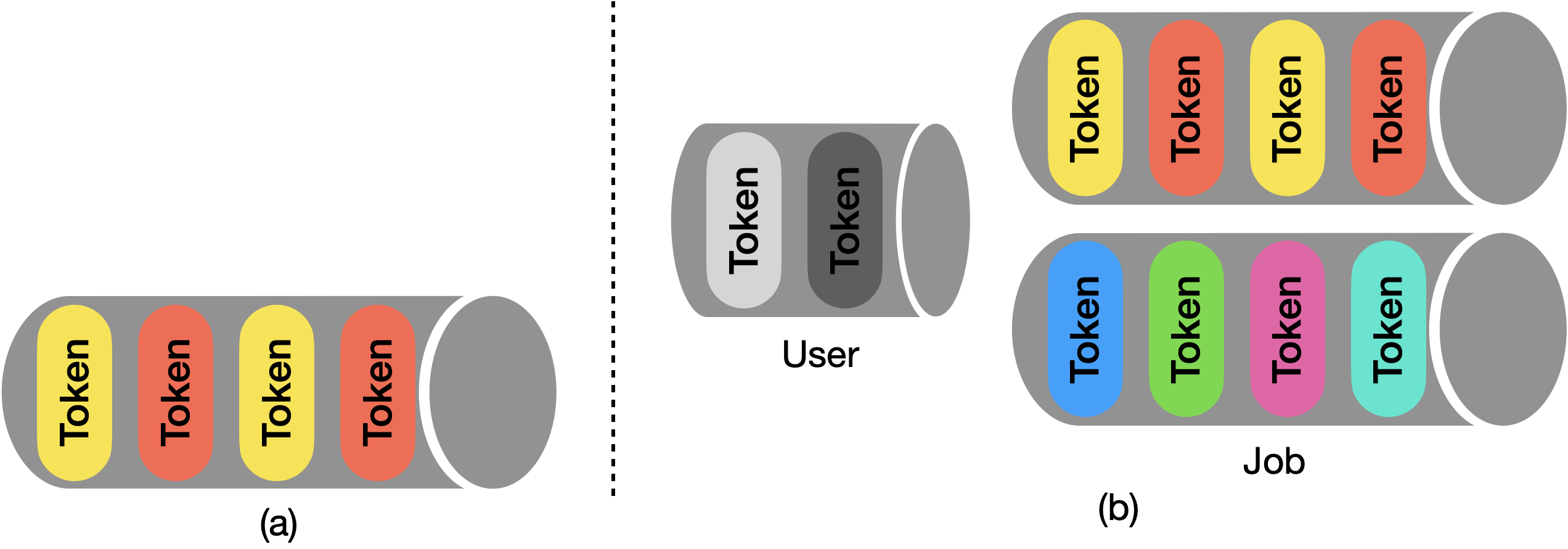}
    \caption{(a) An job-fair token assignment example with two jobs. (b) A user-then-job-fair token assignment of a two-tier token design with two users, one running two jobs and the other running four jobs.  
    \label{fig:token}
}
\end{center}
\end{figure}

Leveraging the classic token mechanism, allocating an appropriate number of tokens to each sharing entity is a straight-forward and effective approach to enable primitive polices, as illustrated in Figure~\ref{fig:token}(a). 
In Figure~\ref{fig:token}(a), we assign two tokens to each job to enable job-fair sharing.
However, this token-based approach is limited in enabling composite policies, e.g., {user-then-job-fair}, where we want to evenly assign I/O resource among users, then enforce job-fair sharing within the scope of a user.
A naive approach to enable this composite policy would to have two tiers of token queues, as shown in ~\ref{fig:token}(b), each representing the sharing entities of users and jobs that belong to a user.
With a composite sharing policy that involves $N$ layers of sharing entity, we have to maintain $N$ tiers of token queues and actual number of token queues is exponential to $N$, which significantly limits the scalability of the token mechanism.
In addition, managing the token queues requires frequent locking and unlocking to ensure consistency, which introduces extra system overhead.

\begin{figure}[h]
\begin{center}
    \includegraphics[width=\columnwidth]{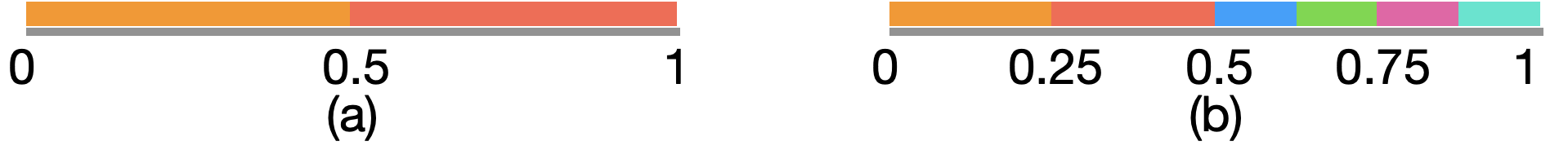}
    \caption{(a) An job-fair statistical token assignment example with two jobs. (b) A user-then-job-fair statistical token assignment of a two-tier token design with two users, one running two jobs and the other running four jobs.  
    \label{fig:prob-token}
}
\end{center}
\end{figure}

We notice that managing the tokens for each sharing entity to enforce a policy is equivalent in a statistical approach:
we divide the range of $[0, 1]$ into several segments, with the segment length proportional to the token counts.
Then an I/O worker draws a random number within $[0, 1]$. 
The I/O request of a job is processed if the random number falls in its corresponding segment.
For primitive policies, the range is split according to the number of jobs, users, or the node counts.
Figure~\ref{fig:prob-token}(a) and (b) show the statistical token assignments of the example in Figure~\ref{fig:token}(a) and (b), respectively.

\begin{figure}[h]
\begin{center}
    \includegraphics[width=\columnwidth]{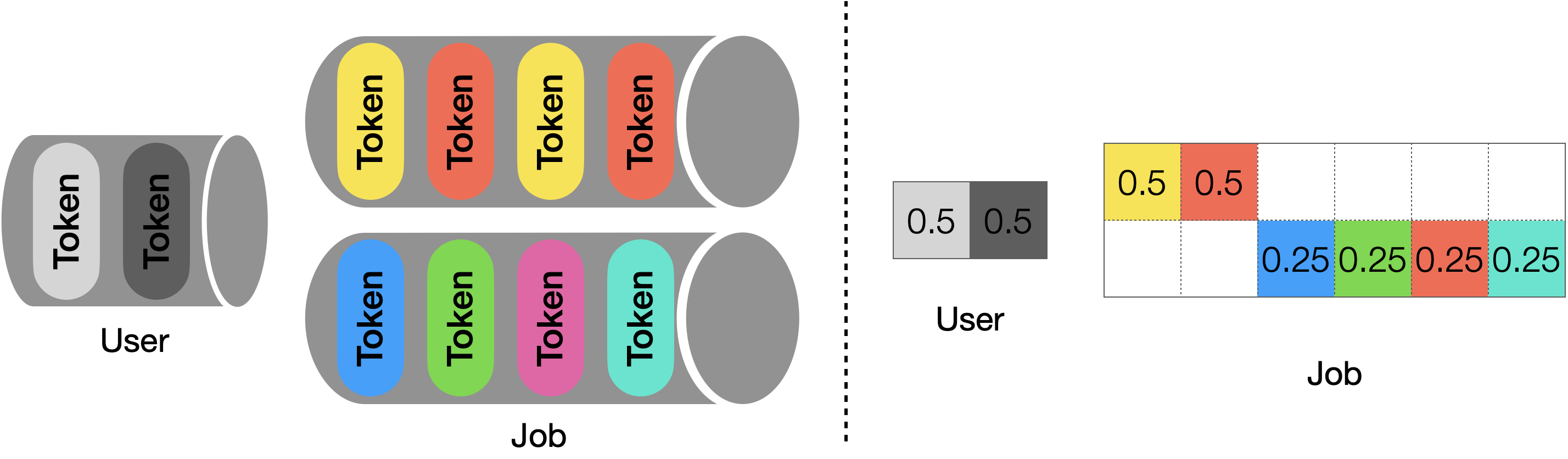}
    \caption{A transition matrix example of the user-then-job-fair policy.  
    \label{fig:transition}
}
\end{center}
\end{figure}

The statistical token assignment can be calculated as a chain of transition matrix multiplication. 
Figure~\ref{fig:transition} shows the transition matrices of the user-then-job-fair example.
At each sharing entity level, each row represents a token queue and each column represent the entities in this level.
For example, in the job matrix of Figure~\ref{fig:transition}, the first row represents the top job queue, where there are two jobs. The second row represents the lower job queue with four jobs.
To derive the statistical token assignment, we multiply the two matrices and obtain the results as in Figure~\ref{fig:prob-token}(b).

More formally, we refer to the transition matrix as $T^i$. The value of $T^i_{j,k}$ is the fair share of the sharing entity as a percentage of its local scope. The sum of each row is one, and only one entry in each column can have a non-zero value, as the sharing percentage is applied within the local sharing entity scope.
The statistical token assignment is evaluated as:
\begin{align}
    \prod_{i=0}^{N-1} T^i, \text{\indent N: the depth of the sharing policy}
\end{align}

Conceptually, this statistical token design is capable of supporting any composite sharing policy with an arbitrary depth.
It reduces the the complex data structure management to a chain production of transition matrices. 
It removes the frequent use of a locking mechanism in synchronized queues at runtime.
The statistical assignment can be easily adjusted by recalculating the matrix multiplication.
One limitation introduced by this approach is that an application has to have a sufficiently large I/O workload to make the statistical token design effective, but this is commonly the case for modern supercomputing applications.

\subsection{Local vs. Global Fairness}
\label{sec:design-fair}
With multiple burst buffer servers, the job information on each server may not always be globally consistent.
If the stripes of every file are spread across all burst buffer servers, e.g., with a sufficiently large stripe number, then every server has the global job status without communication.
Otherwise, if files land on a disjoint set of burst buffer servers, 
every server initially has only local job information, which may not globally consistent.
Synchronization is required to determine global job state and thus, there is a delay before global fairness is reached.

\begin{figure}[h]
\begin{center}
    \includegraphics[width=\columnwidth]{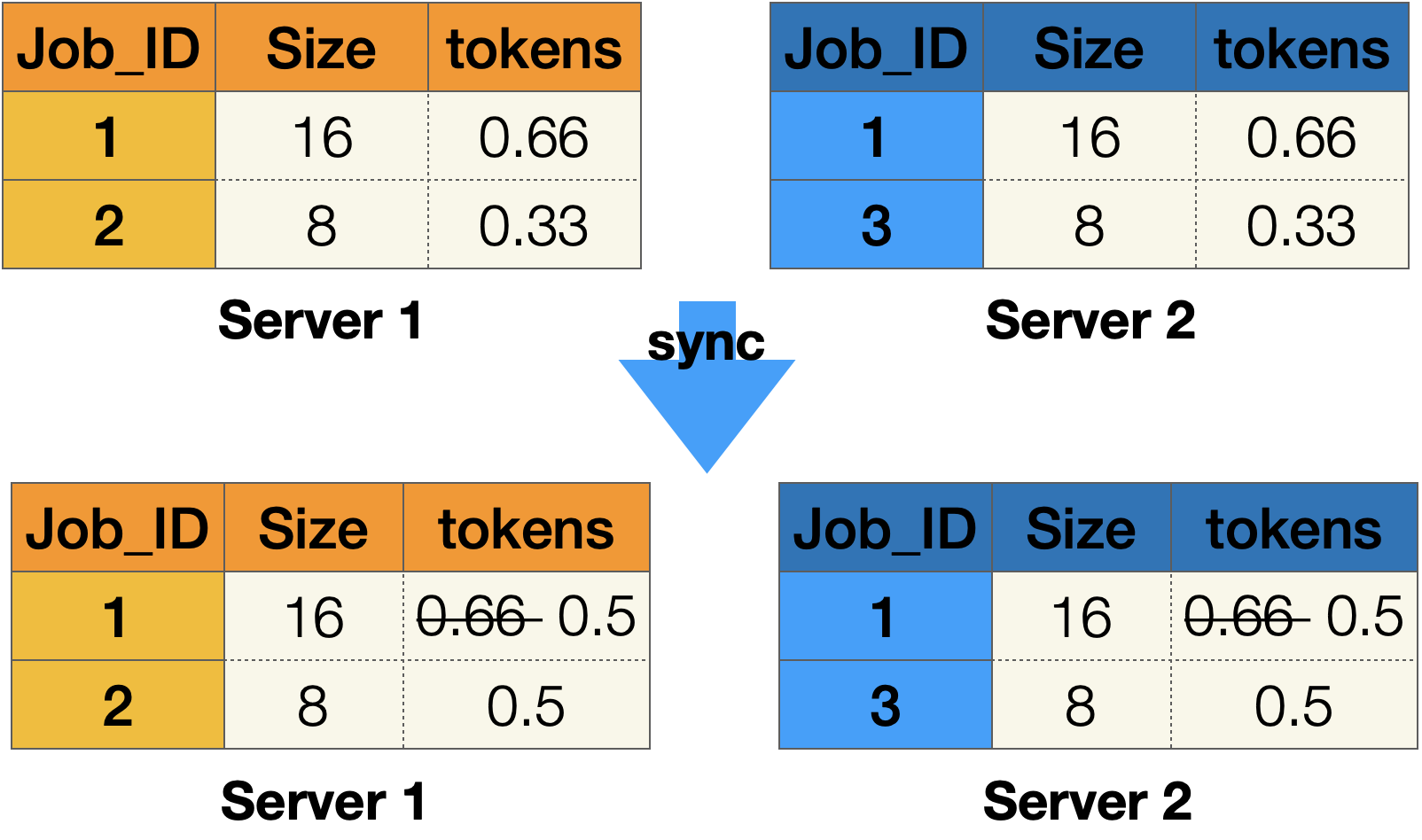}
    \caption{A Simple Example of Job Status Table Synchronization. Server 1 and 2 start with local job status and allocated tokens, then synchronize the tables by exchanging the entries and adding token counts.
    \label{fig:local-global}
}
\end{center}
\end{figure}

Figure~\ref{fig:local-global} shows an example of delayed fairness with \textbf{size-fair}.
Here, Server 1 sees two jobs: Job 1 and Job 2. 
After checking the size of the jobs (16 and 8 compute nodes, respectively), Server 1 assigns $[0, 0.66]$ and $[0.66, 1]$ for the two jobs, respectively.
Similarly, Server 2 assigns $[0, 0.66]$ and $[0.66, 1]$ for Job 1 and Job 3. 
Now Job 1 gets 67\% of the available I/O resources.
However, by looking at the job size globally, we see the correct sharing ratio should be 16:8:8, which means Job 1 should only have 50\% of the I/O resources. 

To address this, \name{} introduces \textbf{$\lambda$-delayed fairness},
where controllers perform an all-gather on the job status table every $\lambda$ time interval. 
This design guarantees that a globally unfair state will not last longer than $\lambda$.
After this communication, both Server 1 and 2 see that Job 1 has a larger range than it should. So every server adjusts the statistical token of Job 1, and global fairness is reached.
An effectiveness study of the length of $\lambda$ is in \S\ref{sec:exp:global}.

\section{Design and Implementation}
\label{sec:design}
In this section, we present the \name{} system's overall design and discuss key components and algorithms that enable generic and global fairness guarantees.

\subsection{Architecture}
\begin{figure}[h]
\begin{center}
    \includegraphics[width=\columnwidth]{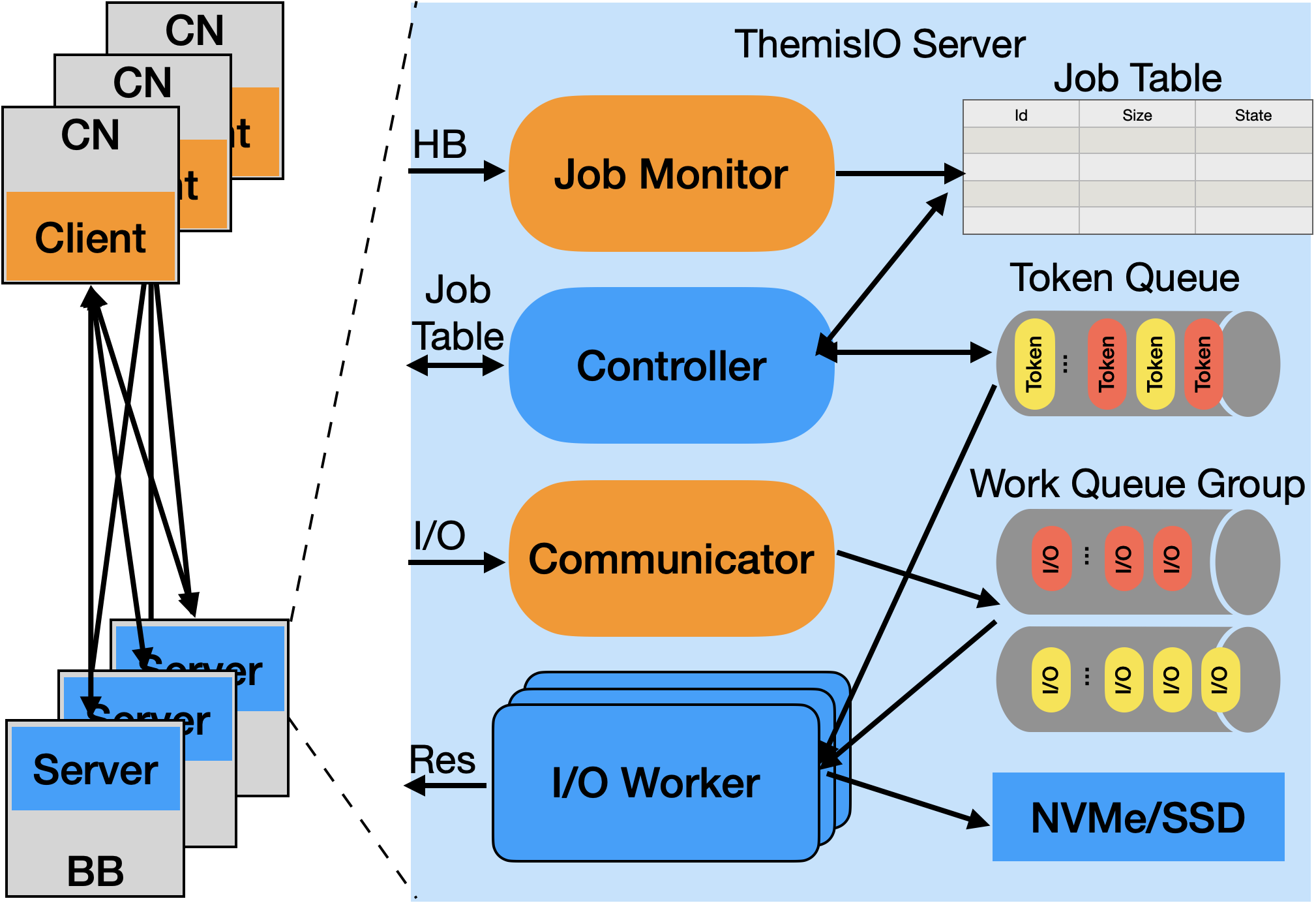}
    \caption{Overview of \name{} Architecture. Clients run on compute nodes with the applications. Servers run on dedicated burst buffer nodes. CN, Compute node; BB, Burst Buffer node; HB, Heartbeat; Res, Results.
    \label{fig:overview}
}
\end{center}
\end{figure}

\name{} exploits a server-client design, as shown in Figure~\ref{fig:overview}. 
{\bf Clients} reside with application processes on compute nodes and {\bf servers} run on the burst buffer nodes.

The client intercepts I/O functions, gathers job metadata 
including user id, job id, and job size, then forwards I/O requests to servers.
The client also sends heartbeats to servers, so that the servers can track job status in real time.

The server has four components: a {\bf job monitor}, an {\bf I/O request communicator}, a {\bf controller}, and a group of {\bf workers}.
The {\bf job monitor} may receive heartbeats from multiple clients of multiple applications. 
It maintains a job status table that summarizes job id, size, user, user group, and status.
Job status is set to active when the corresponding job is new to the server.
It is changed to inactive if a job heartbeat is not received for a predefined period of time.
The {\bf communicator} receives I/O requests from applications. 
Those I/O requests are grouped into queues based on the fair sharing policy.
For example, with size-fair, where I/O resources are proportionally shared with respect to node count, requests are pushed into queues that are identified by job ids.
The {\bf controller} synchronizes with other servers to get the global status of active jobs, and allocates a number of tokens according to the fair sharing policy.
Each {\bf worker} pops one token at a time and an I/O request identified by the token, then processes the I/O request.
There can be multiple workers for higher I/O throughput.

\subsection{Communication}
\name{} uses Unified Communication X (UCX~\cite{shamis2015ucx}) for low latency and high bandwidth communication. Each \name{} server maintains two types of UCP workers: one for client-server communication and the other for server-server communication. A UCP worker is an opaque object in UCX that represents an instance of a local communication resource and the progress engine associated with it. When a \name{} server accepts a connection request from a \name{} client, it assigns a UCP worker to the client and keeps a mapping from the client to the UCP worker. One UCP worker can be shared between multiple \name{} clients.

Each application process/thread launches a \name{} client, which initializes a local UCP worker with a \name{} server. When the endpoint connection is established on the local UCP worker for the destination \name{} server address, job metadata is transferred to the servers. I/O requests are forwarded to the \name{} servers via the UCP worker to be processed, and the results are returned to the \name{} client. The heartbeat monitor in \name{} servers monitors the job status, and if a job is inactive for a predefined period of time, the server marks the job's status as inactive and destroying all the UCP worker resources mapping entries associated with that job. When a client exits, it notifies the \name{} servers to destroy the corresponding mapping entry.

The UCP workers are persistent during the lifetime of \name{}.
%\katznote{I don't know that we need to talk about it here, but what happens if a server crashes? This leads me to wonder about fault tolerance of all the components more generally.} 
%\zhao{We haven't implemented the fault tolerance part yet, as we focus on sharing. For now, if application or client crashes, then users can relaunch the job. If the server crashes, we don't have a good way of handling failure. I had a plan to look into log structure file system on SSD, but we haven't implemented it yet.}
%\katznote{do we need to say anything here, or in Future Work/Conclusions?}
%\zhao{sure, I added one sentence in Future Work.}
\name{} servers synchronize the job status table to have a global view of the jobs and the assigned tokens.
This synchronization enables the controller to adjust the token count to achieve $\lambda$-fairness, as discussed in \S\ref{sec:design-fair}.

\subsection{File System}
\label{sec:impl-fs}
We integrate \name{} with a byte-addressable file system to support NVMe or SSD. 
To gain complete control and native speed, we implement a user space file system ourselves, however, \name{} can work with any shared file system in either kernel or user space.
%\katznote{do we want to say that this is for our experiments? How would we work with a production file system}
%\zhao{How about this: ``To gain complete control and native speed, we implement a user space file system for our experiments, however, \name{} can work with any shared file system in either kernel or user space. ''}
%
In this file system, both directories and files are stored as files,
and files and metadata are spread across \name{} servers using a consistent hash function.
Striping is supported with corresponding records in file metadata.

%A file is located by using a consistent hash function then with the stored address in NVMe on the hosting server.
The location of a file is on one or more servers, determined by a hash function, and on those servers, an index specifies the NVMe region of the file's contents.
Directory and file creation updates the content of the parent directory.
Queries over a directory return the content in that directory.
Reading a file returns the contents specified by the path and offset range. 
Writing a file writes/overwrites a range of allocated byte-addressable space in NVMe,
and the metadata update is done on the same \name{} server.
Concurrent read operations on the same file are executed without locking. 
Concurrent write operations to the same file proceed without any limitation if the byte ranges do not conflict.
However, a locking mechanism is used when multiple threads are updating the file metadata.

\subsection{I/O Function Interception}
One of the design principles of \name{} is to be compatible with existing applications, i.e., applications do not have to make code change to leverage \name{}.
However, most production supercomputers do not grant root privilege, which makes a kernel module implementation infeasible.
So \name{} uses a I/O function interception technique.
In this way, \name{} provides a POSIX-compliant interface, with which users can simply point I/O to a path that is prefixed by \name{} namespace, e.g., \texttt{/fs/input/path}.
All I/O to/from this path will be intercepted by \name{} then processed in burst buffers.

To implement this, we either intercept the 64-bit version of I/O functions in the GNU C Library by simply exposing functions with the same name, or rewrite the first several instructions of a function with a jump instruction to the function implemented in \name{} library, then jump back to the original function if necessary.
The first method is referred to as \textbf{override}~\cite{kerrisk2014linux} and the second method as \textbf{trampoline}~\cite{hunt1999detours}. 

Listing~\ref{lst:interface} summarizes the intercepted functions.

\begin{lstlisting}[style=CStyle, caption={Example Intercepted Functions}, label={lst:interface}]
int open(const char *filename, int flags[, mode_t mode])
int close(int fd)
ssize_t read(int fd, void *buffer, size_t size)
ssize_t write(int fd, const void *buffer, size_t size)
off_t lseek(int fd, off_t offset, int whence)
DIR * opendir (const char *dirname)
struct dirent * readdir(DIR *dirstream)
int closedir (DIR *dirstream)
int stat(const char *filename, struct stat *buf)
\end{lstlisting}

\section{Experiments and Results}
\label{sec:experiments}
To validate the effectiveness of the \name{} design and the correctness of our implementation, we run both benchmark and real applications with various sharing policies. 
We also implement the sharing algorithms in GIFT and TBF in \name{} and perform a comparative study.
In addition, we investigate the impact of system performance with various communication intervals in $\lambda$-delayed fairness.
In summary: 
\begin{itemize}
\item{} Benchmarks show that \name{} can efficiently share I/O resources between jobs with both primitive and composite policies (details in \S\ref{sec:exp:policy}).
\item{} The comparison study with FIFO, GIFT, and TBF shows that \name{} shares I/O resources more efficiently and stably (details in \S\ref{sec:exp:compare}).
\item{} Real applications show that \name{} reduces the I/O intervention slowdown drastically or completely (details in \S\ref{sec:exp:apps}).
\item{} The communication interval in $\lambda$-delayed fairness can be as large as 500~ms without significant impact on global fairness. (details in \S\ref{sec:exp:global})
\end{itemize}

All the experiments are run on the TACC Frontera supercomputer, which consists of 8,008 CPU nodes (CLX), 16 large-memory nodes (NVDIMM), and 90 GPU nodes (RTX).
The CPU nodes have two Intel Xeon Platinum 8280 processors and 192~GB RAM.
Each large-memory node has four Intel Xeon Platinum 8280 processors and 2.1~TB RAM, supported by Intel Optane memory.
Each GPU node has four Nvidia Quadro RTX 5000 cards and 128~GB RAM.
In all experiments, \name{} runs on the CLX nodes with RAM as storage devices. 

\subsection{Benchmark and Application Configuration}
Throughout the experiments, we used IOR and mdtest for simple tests and a customized benchmark to measure I/O sharing performance.
The customized benchmark simulates two workloads:
1) {\it iops\_stat} repeatedly calls stat() to query file metadata with randomly generated file names;
2) {\it iops\_write\_read} writes a small (1~MB) file then reads the same file repeatedly.
We disable client caching in all tests as \name{} is designed for remote-shared burst buffer, and we are investigating the I/O sharing capability in particular.

To simplify validation of the sharing capability, we run each application at a fixed size. Some of the applications take too long to finish, so we only run cases with a reasonable time length, which are still representative of their I/O workloads.
The \textbf{WRF} benchmark uses the 12 KM CONUS Benchmark dataset from \url{https://www2.mmm.ucar.edu/wrf/WG2/benchv3/#_Toc212961288}. It is a 48-hour, 12-km resolution case over the Continental U.S. (CONUS) domain October 24, 2001 with a time step of 72 seconds. We run the WRF benchmark on 4 nodes each with 56 MPI process per node. 
The \textbf{NAMD} benchmark uses the 1M atom Satellite Tobacco Mosaic Virus system from \url{https://www.ks.uiuc.edu/Research/namd/benchmarks/}. It runs on 64 nodes with 8 MPI processes per node and 7 threads per process. The input was modified to save trajectory every 48 steps. 
The \textbf{SPECFEM3D} benchmark runs a small-scale regional seismic wave propagation simulation tweaked from the benchmark data set published by NVIDIA (\url{https://www.nvidia.com/en-sg/data-center/gpu-accelerated-applications/specfem3d-globe/}). 
The grid is defined in one cubed-sphere chunk of the globe and sliced into 224x256 elements. 
The simulated record length is 100 minutes. 
We run the benchmark on 16 nodes with 56 MPI processes per node.
The \textbf{ResNet-50} case uses an open source PyTorch implementation\footnote{anonymized for peer review} with the ImageNet~\cite{deng2009imagenet} dataset that contains 1,331,167 images.
The total size is $\sim$156~GB and the average size of the image is about 116~KB. 
We run ResNet-50 on 16 RTX nodes with a 128 batch size per GPU. 
The complete training time is $\sim$20 hours, so we only use the first three epochs ($3\times157$ steps) in this case. 
The \textbf{BERT} case is a PyTorch implementation\footnote{anonymized for peer review} with the English wikitext and Toronto Book Corpus datasets.
The text is reorganized as 512 HDF5 files, with a total size of 71~GB and an average size of $\sim$48~MB. 
We run BERT on four RTX nodes with a 16 batch size per GPU.
BERT training has two phases, where phase 1 takes 393~hours to finish on four RTX nodes, so we only use the first three steps in this case.

\subsection{Scaling Performance}
\label{sec:exp:scaling}
\begin{figure}[h]
\begin{center}\up
    \includegraphics[width=\columnwidth]{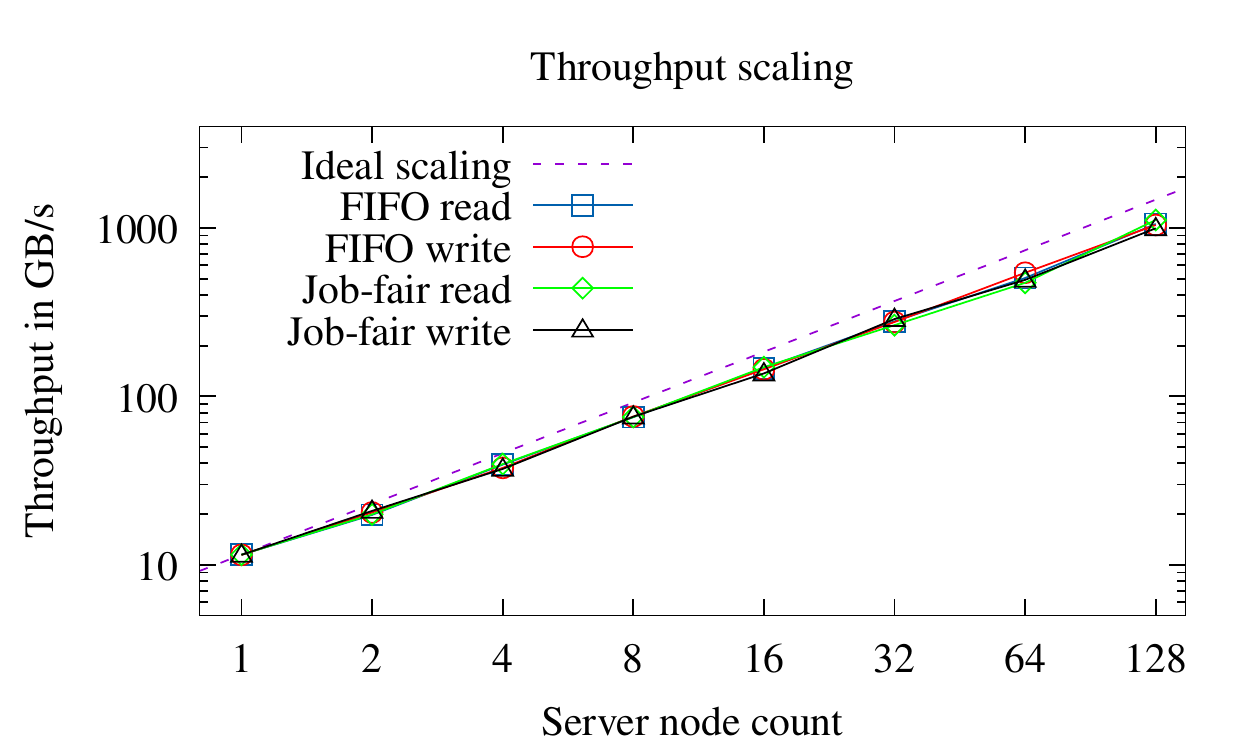}
    \caption{Aggregate throughput when using multiple \name{} server nodes
    \label{fig:scaling_ior_throughput}
    }
\end{center}
\end{figure}

Figure~\ref{fig:scaling_ior_throughput} represents the unidirectional aggregate throughput achieved running the \name{} server on 1 to 128 nodes. For each set of server nodes, an equal number of nodes were each running eight IOR processes, writing and reading 1 GB files in 1 MB blocks. We include a comparison of the performance of simple FIFO queuing versus job-fair queuing. With one server node, this achieved a maximum throughput of 11.7 GB/s. With eight server nodes, the slowest result was for FIFO reads at 77.1 GB/s, a scaling efficiency of 82\%.
With 128 server nodes, the throughput reached 1017 GB/s, a scaling efficiency of 68\%. 
For comparison, the largest Lustre file system on Frontera has 32 OST nodes with an aggregate throughput of 120 GB/s.
The DataWarp burst buffer on NERSC Cori has a peak throughput of 1.7~TB/s with each burst buffer node contributing 6.5~GB/s.
The sustained I/O throughput of \name{} is comparable to the state of the art production system.
It is worth noting that these experiments were unidirectional, just writing or just reading, unlike the read/write tests in Figure~\ref{fig:bench-share}, where only half the interconnect throughput is available.

\subsection{Sharing with Various Policies}
\label{sec:exp:policy}
In this section, we present \name{}'s sharing capability with various primitive and composite policies.
%Then we compare the sharing efficiency of \name{} with existing algorithms of GIFT and TBF.

\subsubsection{Primitive Policy}
\label{sec:exp:policy:prim}
\begin{figure*}[t]
  \subfigcapmargin=0.1in
  \subfigure[Size-fair, 4-node job competing with 1-node job]{
    \includegraphics[width=2.25in,clip]{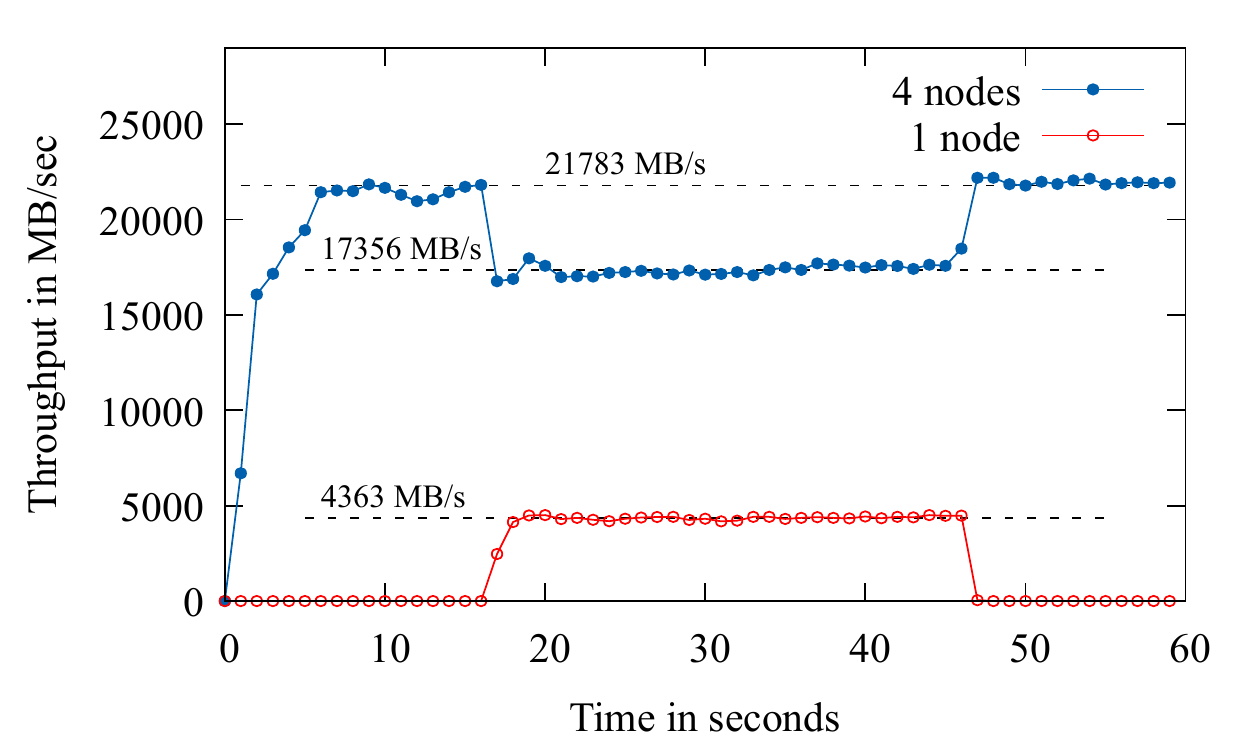}%
    \label{fig:size-fair}
  }
  \subfigure[Job-fair, 4-node job competing with 1-node job]{
    \includegraphics[width=2.25in,clip]{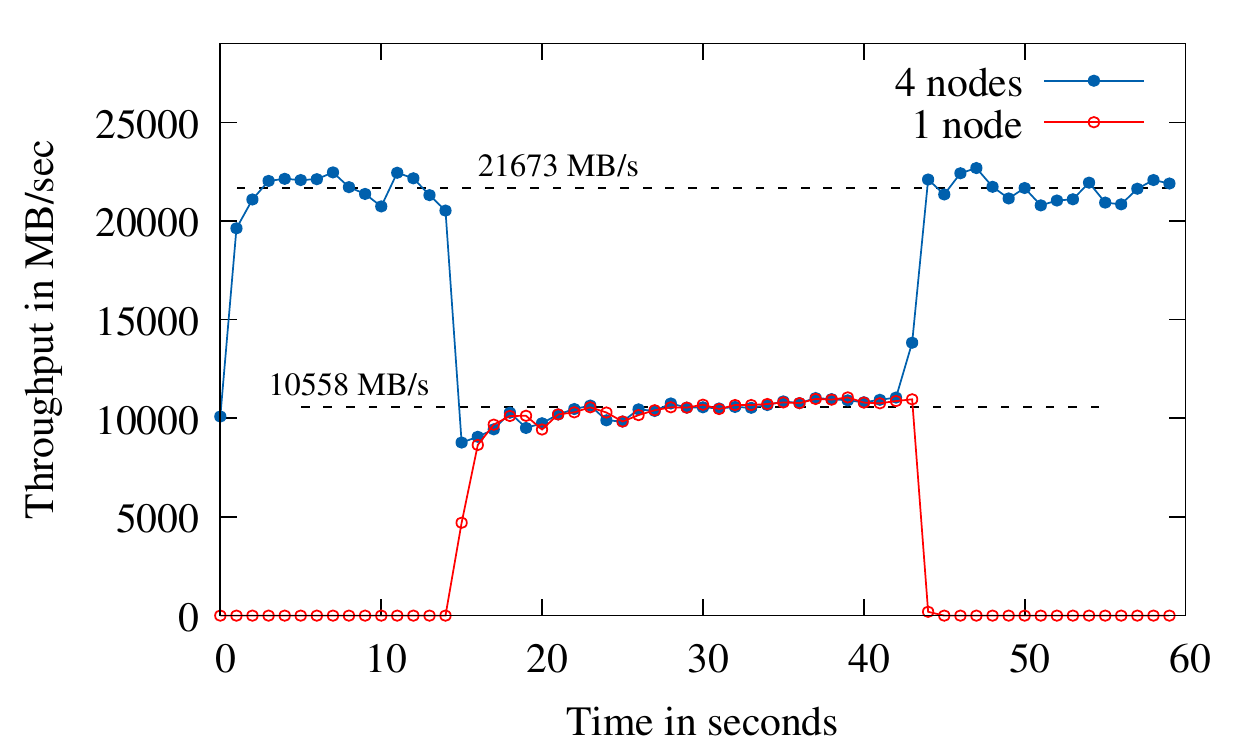}%
    \label{fig:job-fair}
  }
  \subfigure[User-fair, Two 2-node jobs competing with a 1-node job]{
    \includegraphics[width=2.25in,clip]{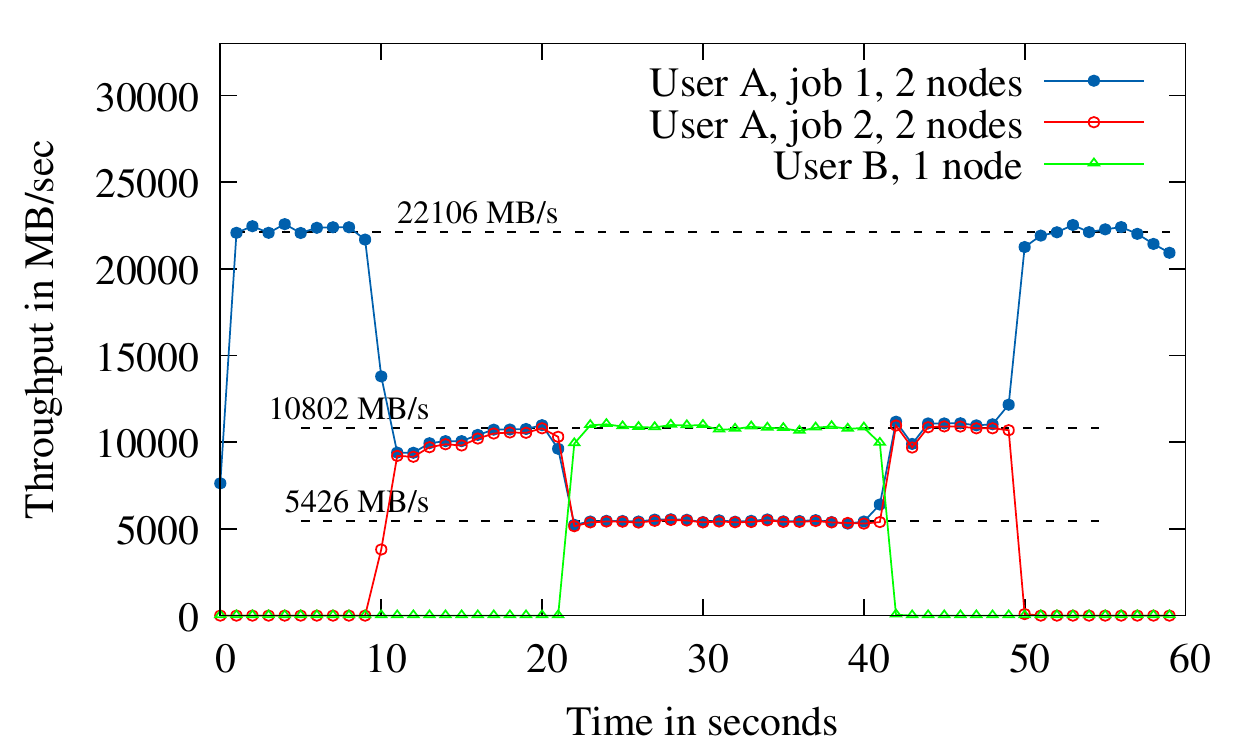}%
    \label{fig:user-fair}
  }
  \caption{Effectiveness of I/O Resource Sharing with Size-, Job-, and User-fair Policies with Single \name{} Server.}
  \label{fig:bench-share}\up\up\up
\end{figure*}

Figure~\ref{fig:bench-share} shows the results of three primitive sharing policies, namely, {\bf size-fair}, {\bf job-fair}, and {\bf user-fair}.
We concurrently run two benchmark programs on different numbers of nodes and report the throughput (MB/s) to examine the sharing efficacy. The benchmark program in these experiments opens one file per process. Each process writes 10~MB of data to its file, then reads it back, and continues to repeat this write/read cycle for a set length of time. 
Figure~\ref{fig:bench-share} demonstrates the measured I/O throughput with samples taken at 1-second intervals. 
The actual response time of each I/O operation is on the order of 1 microsecond. 
The two-second delay is a measurement artifact.

Figure~\ref{fig:size-fair} demonstrates a server running in {\bf size-fair} mode with a 224-process benchmark job running on four nodes competing for server throughput with a benchmark job consisting of 56 processes running on one node. The first job runs for 60 seconds, while the second job run for 30 seconds, starting 15 seconds after the first job starts. The median throughput of the 4-node job when running unopposed is 21.8 GB/sec. The median throughput of the 4-node job and the 1-node job when both are running is 17.4 GB/sec and 4.4 GB/sec, respectively. This represents a throughput ratio of 3.96x, which closely approximates the 4x ratio of job sizes.

Figure~\ref{fig:job-fair} demonstrates the same pair of benchmark jobs, but with a server running in {\bf job-fair} mode. The first job still consists for four times as many client processes, but the overall throughput for the jobs when both are running is nearly equal, with a median throughput of 10.6 GB/sec.

In Figure~\ref{fig:user-fair}, the server is configured in {\bf user-fair} mode, and three jobs from two users compete for throughput. User A is running two jobs, each of which uses two nodes, while User B is running one job on one node. When all three jobs are running, User A's jobs have a median total throughput of 10.85 GB/sec, which is roughly the same as the throughput of User B's job: 10.80 GB/sec.

~\zhao{add one paragraph about ThemisIO utilization. Add a ref in intro.}

\subsubsection{Composite Policy}
\label{sec:exp:policy:com}
Next, we examine the I/O sharing efficiency with composite policies.
In particular, we choose the {\bf user-then-size-fair} and {\bf group-user-then-size-fair} policies. 

\begin{figure}[ht]
  \up\up\up
  \includegraphics[width=\columnwidth]{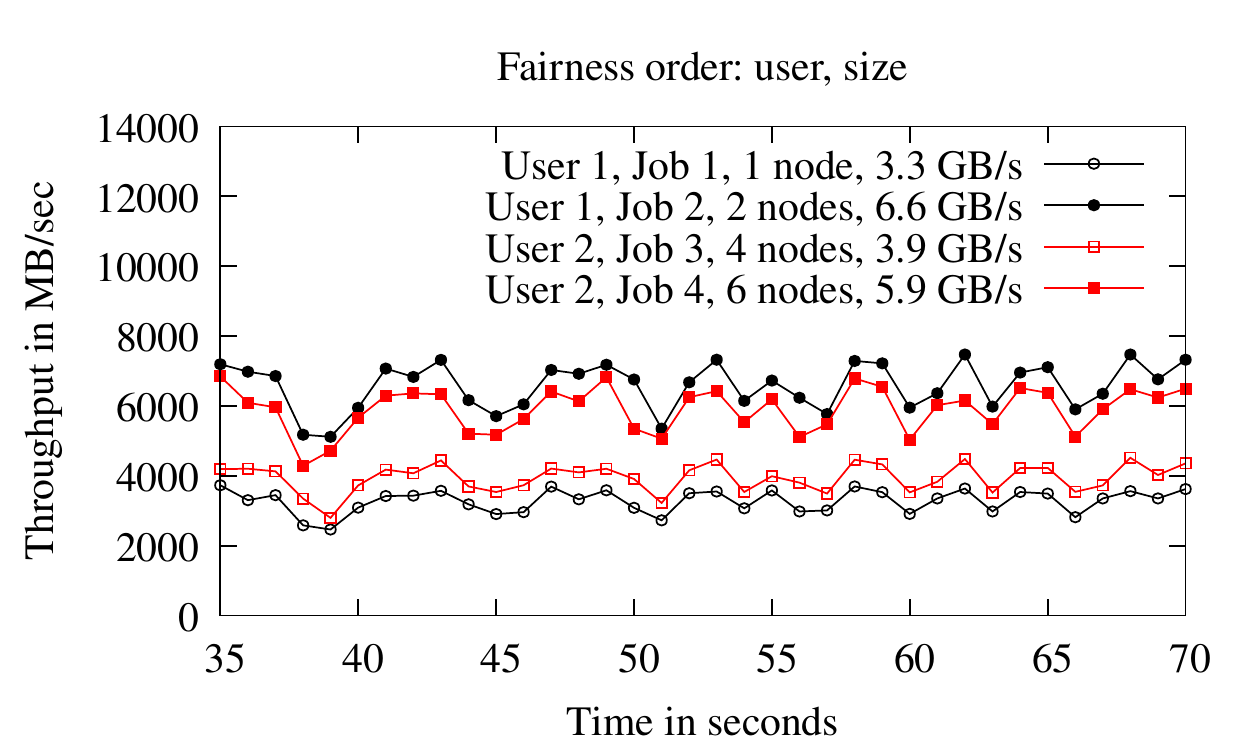}
  \caption{Four competing jobs, two from each user. Total throughput is balanced between users and between jobs belonging to each user.}
  \label{fig:user-size-fair}
  \up
\end{figure}

In the first experiment, we run four jobs owned by two users with different node count. 
As shown in Figure~\ref{fig:user-size-fair}, user-fair sharing is achieved at the first level, as the two jobs of user 1 get a throughput of 10.1~GB/s, while user 2 gets 9.9~GB/s. 
Looking deeper into the two jobs of user 1, Job 1 gets 3.4~GB/s and Job 2 gets 6.7~GB/s, which matches the ratio between node count of 1:2.
Similarly, Job 3 and 4 get 3.9~GB/s and 6.0~GB/s, respectively.
It is close to the node count ratio of 4:6.
The aggregated throughput is 20~GB/s, which is $\sim$1.7~GB/s lower than the primitive policy.
This throughput degradation is caused by the slow startup of \name{} when setting up the connections between clients and servers, which is negligible compared to the long runtime of modern supercomputing applications.

\begin{figure}[ht]
  \up\up\up
  \includegraphics[width=\columnwidth]{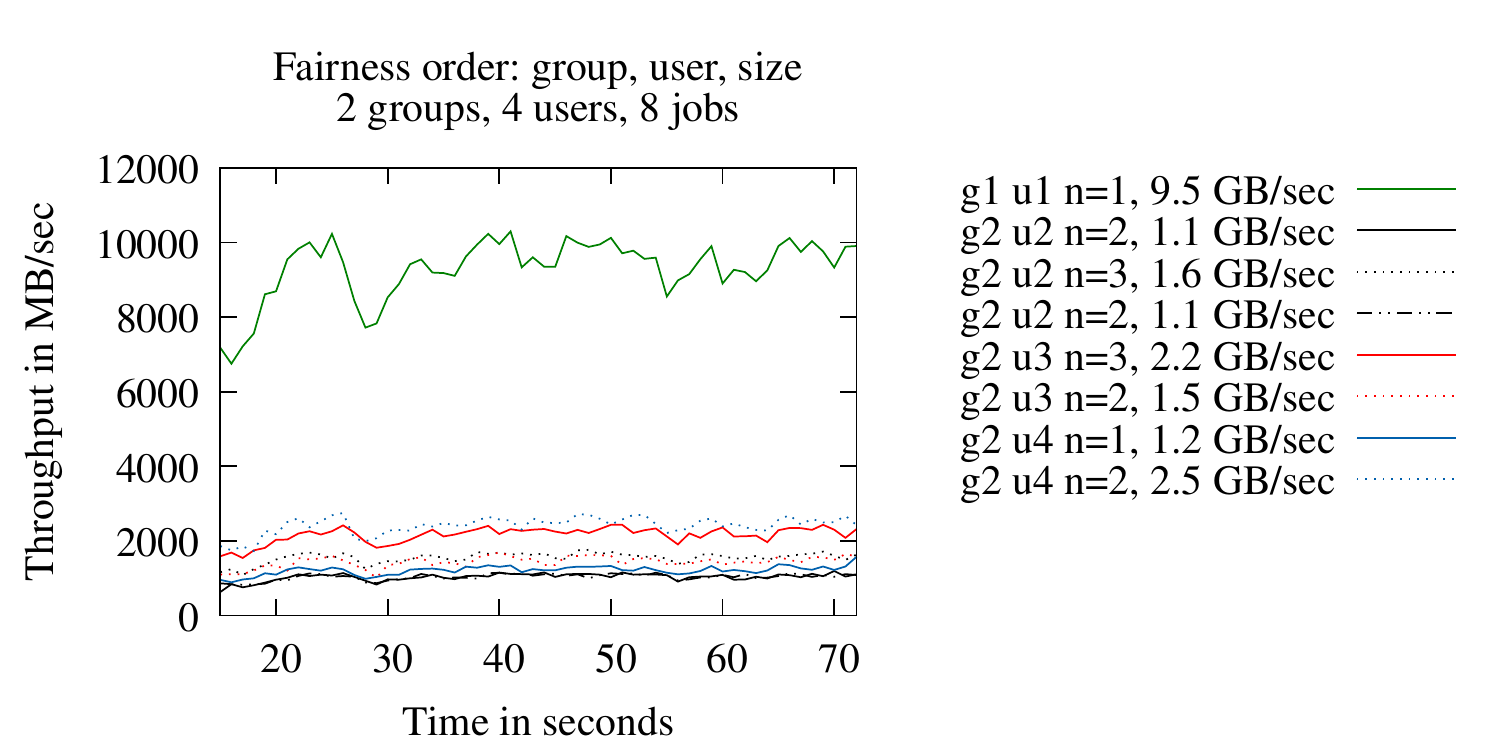}
  \caption{Eight competing jobs from two groups containing four users. Throughput is balanced by group, then by user, then by size.}
  \label{fig:group-radical}
\end{figure}

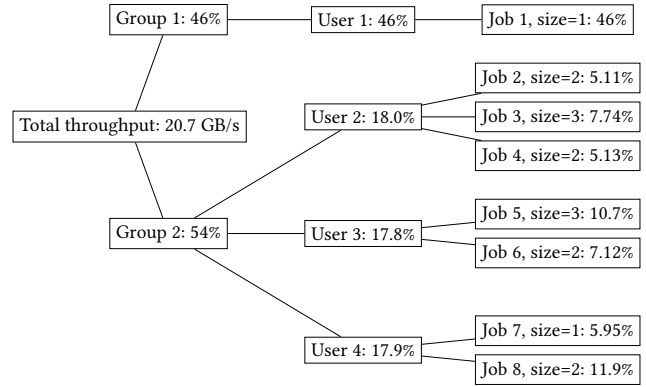
\begin{figure}[h]
\resizebox{\columnwidth}{!}{
\begin{tikzpicture}
[sibling distance=10em,
  every node/.style={shape=rectangle,draw,align=center},
  grow=right]
\node{Total throughput: 20.7 GB/s}
  [sibling distance=11em,level distance=2em]
child{
  node{Group 2: 54\%}
  [sibling distance=6em,level distance=10em]
  child{
    node{User 4: 17.9\%}
    [sibling distance=2em, level distance=10em]
    child{node{Job 8, size=2: 11.9\%}}
    child{node{Job 7, size=1: 5.95\%}}
  }
  child{
    node{User 3: 17.8\%}
    [sibling distance=2em, level distance=10em]
    child{node{Job 6, size=2: 7.12\%}}
    child{node{Job 5, size=3: 10.7\%}}
  }
  child{
    node{User 2: 18.0\%}
    [sibling distance=2em, level distance=10em]
    child{node{Job 4, size=2: 5.13\%}}
    child{node{Job 3, size=3: 7.74\%}}
    child{node{Job 2, size=2: 5.11\%}}
  }
}
child{
  node{Group 1: 46\%}
  [level distance=10em]
  child{
    node{User 1: 46\%}
    [level distance=10em]
    child{
      node{Job 1, size=1: 46\%}
    }
  }
}
;
\end{tikzpicture}
}
  \caption{Tree depiction of group-user-size-fair policy experiment. Each node lists the percentage of the overall throughput allocated to that job, user, or group of users. Note that the throughput is approximately balanced across groups and users within a group, and proportional to job size across jobs for each user.}
  \label{fig:tree-group-user-size}\up\up
\end{figure}

To demonstrate the flexibility of \name{} in supporting composite policies, we also implement a three-tier {\bf group-user-size-fair} policy. 
This policy should enforce an even I/O throughput partition across groups, then across users in each group, allocating the I/O resource among jobs of each user in proportion to the job size.
Figure~\ref{fig:group-radical} shows the results of the experiment with two groups, four users, and eight jobs,
and Figure~\ref{fig:tree-group-user-size} shows the results as a hierarchy tree.
Group 1 gets 9.5~GB/s and Group 2 gets a total of 11.2~GB/s.
This is not an exactly even split due to the slow startup in first 10 seconds, as in the previous experiment.
However, I/O resource are almost fair-shared after the startup period.
Inside Group2, User 2, 3, and 4 get total throughput of 3.8~GB/s, 3.7~GB/s, and 3.7~GB/s, respectively.
I/O through is evenly split between the three users.
For each user, all jobs get throughput proportional to the job size. 
That is to say, the three jobs of User 2 get 1.1~GB/s, 1.6~GB/s, and 1.1~GB/s, which is almost the ratio of 2:3:2.
The overall throughput is 20.7~GB/s, which is only 1~GB/s lower than the maximum throughput.

In summary, \name{} can effectively and efficiently support primitive policies that are as easy as size-fair, job-fair, and user-fair.
It is also capable of supporting composite polices such as user-size-fair and group-user-size-fair.

\subsection{Comparison with Existing Solutions}
\label{sec:exp:compare}
In this experiment, we compare the I/O resource sharing performance of \name{} with existing solutions of GIFT and TBF using the metrics of overall I/O throughput, latency to fair-sharing, and the standard deviation of I/O throughput. 

To understand the performance of \name{} vs. that of GIFT, we copy the GIFT core algorithms, BSIP (Basic Synchronous I/O Progress) and the linear programming algorithm, from the GIFT codebase into \name{} and replace the I/O resource allocation and throttling mechanisms of LINUX cgroups with the \name{} probabilistic token design. 
%\katznote{It's not clear to me why this is a fair comparison.}
%\zhao{The original GIFT is implemented with cgroups to apply I/O throuttling, which is a coarser grained solution. To just compare the overhead of the sharing algorithms, we reimplement the same algorithms using the same ThemisIO framework.}
%\katznote{how do we know we do this efficiently, or as well as the GIFT team would do this?}
%\zhao{We copy \& paste the BSIP and LP algorithms from the GIFT code base. These two algorithms decide the I/O resource allocation, and we integrate these algorithms with ThemisIO statistical token.}
%\katznote{you've said what we do, but }
GIFT uses pending I/O requests every $\mu$ time interval to determine bandwidth allocation. 
The default setting of $\mu$ is ten seconds, which leads to a long delay in I/O resource adjustment.
We experiment with a series of $\mu$ values and conclude that 0.5 sec is an appropriate interval for our reference implementation.
The original implementation of TBF on Lustre directly manages tokens, which are assigned based on I/O request type.
Similarly to GIFT, we implement the core HTC (Hard Token Compensation) and PSSB (Proportional Sharing Spare Bandwidth) strategies and integrate them with \name{}'s I/O resource allocation mechanism.

\begin{figure*}[]
  \subfigcapmargin=0.1in
  \subfigure[Job-fair Sharing with \name{}.]{
    \includegraphics[width=2.2in,clip]{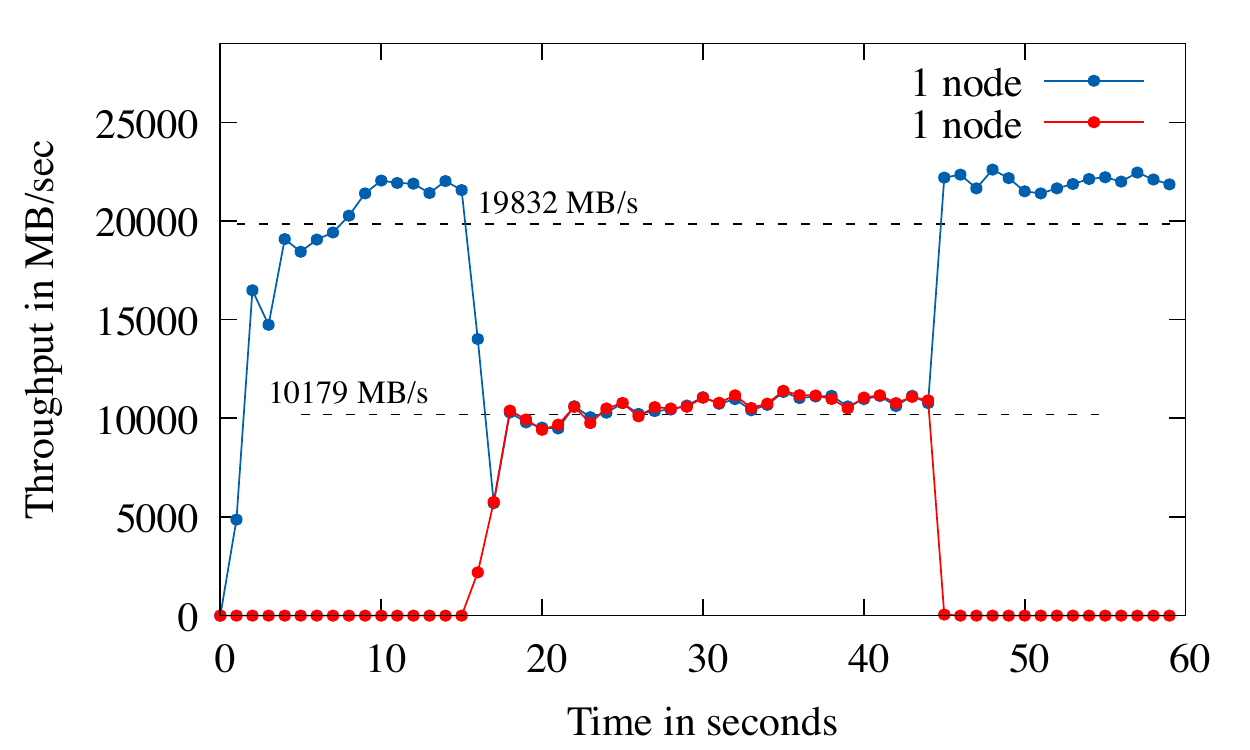}
    \label{fig:compare-themisio}
  }
  \subfigure[Job-fair Sharing with GIFT.]{
    \includegraphics[width=2.25in,clip]{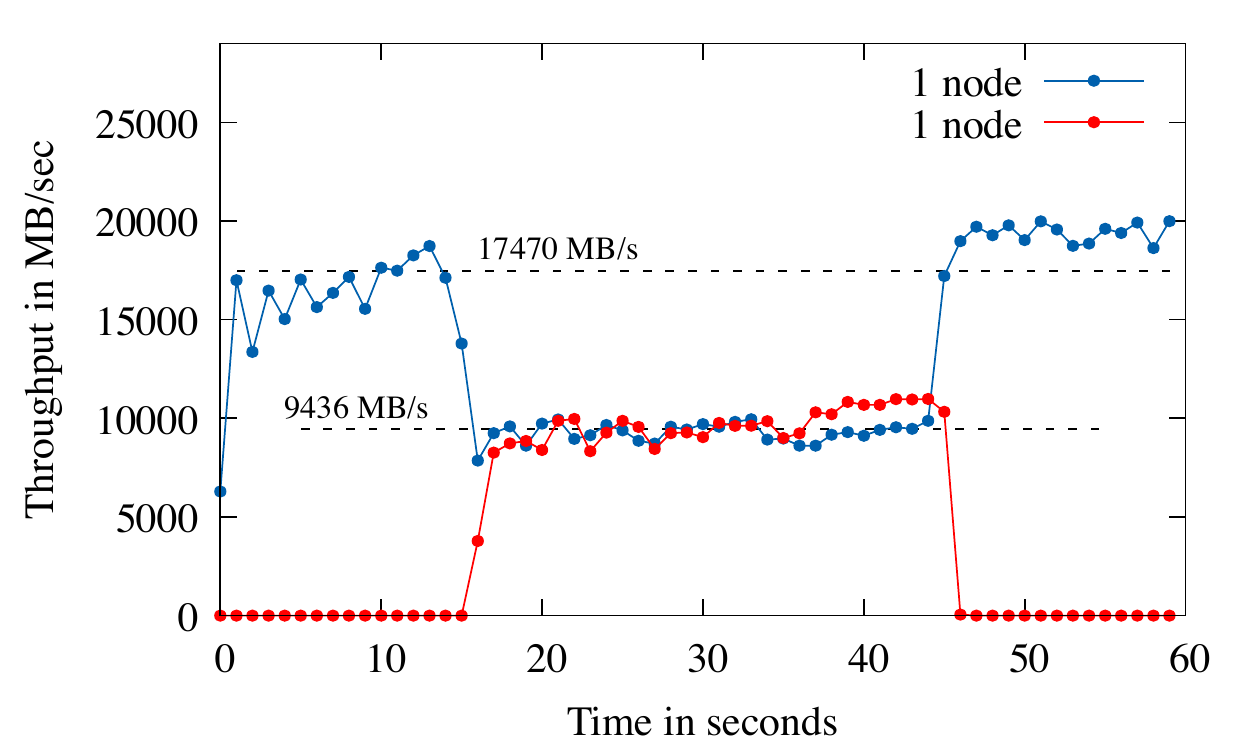}
    \label{fig:compare-GIFT}
  }
  \subfigure[Job-fair Sharing with TBF.]{
    \includegraphics[width=2.2in,clip]{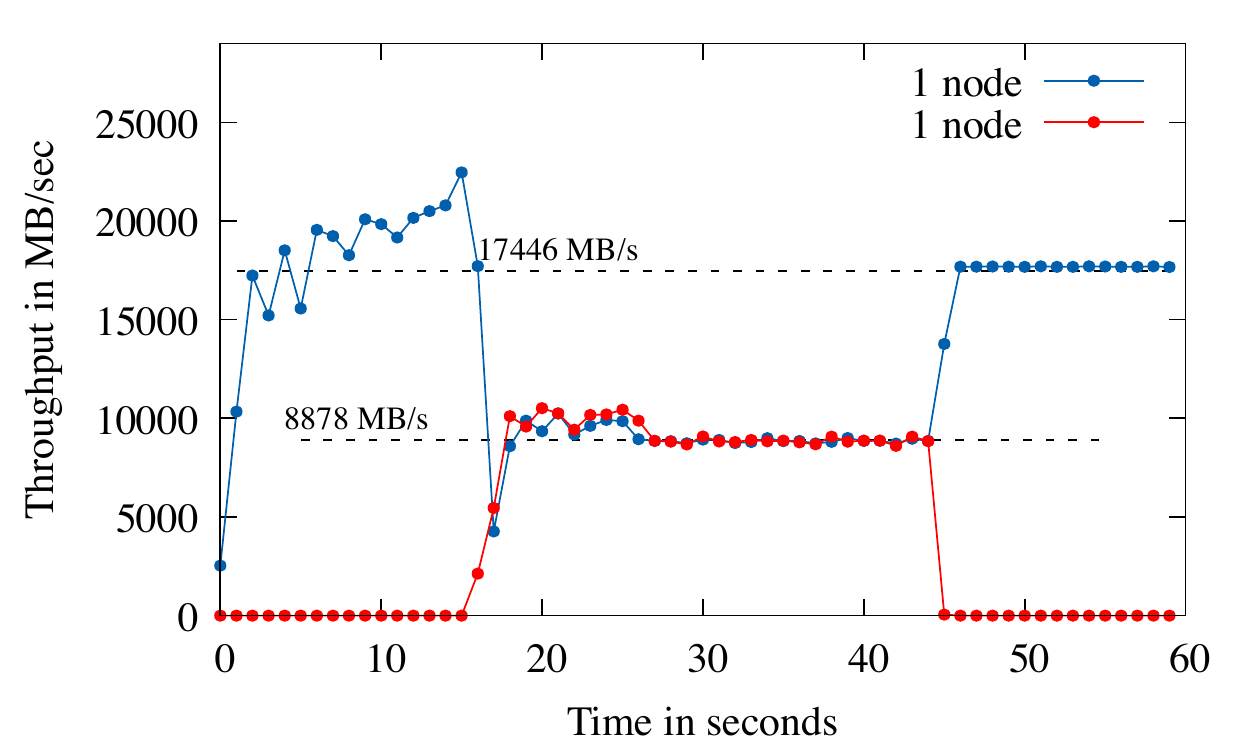}
    \label{fig:compare-TBF}
  }
  \caption{I/O Resource Sharing Comparison with Existing Solutions with Single \name{} Server.}
  \label{fig:comparison}
\end{figure*}

Figure~\ref{fig:comparison} presents the comparison of \name{} with GIFT and TBF using a pair of single node benchmark jobs.
In each experiment, Job 1 runs for 60 seconds and Job 2 is started 15 seconds after Job 1 and runs for 30 seconds.
\name{} runs in job-fair mode.
The sustained peak throughput of \name{} is 19.8 GB/s, which is 13.5\% and 13.7\% higher than that of GIFT and TBF, respectively.
During the sharing phase, the throughput of Job 2 is 10.2 GB/s, which is 7.9\% and 14.7\% higher than GIFT and TBF.
\name{} also shows a lower standard deviation of the throughput of Job 2 with a value of 504~MB/s, compared with 626~MB/s for GIFT and 845~MB/s for TBF.
Compared with existing solutions, \name{} shares I/O resources more efficiently and more stably. 
In addition, GIFT and TBF only support job-fair sharing and require prior knowledge, e.g., the repeated pattern and the I/O rate.
In contrast, \name{} is more versatile in sharing policies and it gathers all necessary information at runtime from the I/O traffic, which makes I/O resource sharing adaptive to the real workload without requiring user-supplied information.

\subsection{Sharing with Applications}
\label{sec:exp:apps}
\begin{figure}[t]
\begin{center}
    \includegraphics[width=83mm]{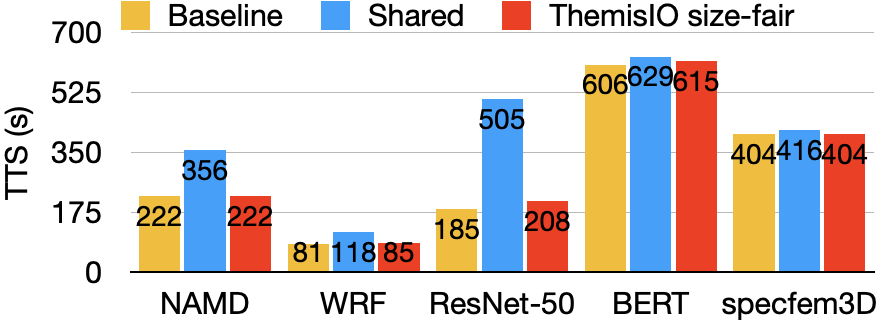}
    \caption{Relative FIFO and size-fair time-to-solution to the baseline with exclusive \name{} access. This experiment uses two \name{} servers.}
    \label{fig:apps}\up\up\up\up
\end{center}
\end{figure}

In previous experiments, we showed that \name{} can correctly and efficiently share I/O resources among jobs with benchmarks.
In this experiment, we study the overall impact of policy-driven I/O resource sharing on applications.
We run the five applications 1) with exclusive access to a \name{} deployment of one server except ResNet-50 which runs with two servers due to space limit, 2) with FIFO policy and a background benchmark job (one compute node), and 3) with size-fair policy and the background benchmark job. 
We expect that with {\bf size-fair}, \name{} can significantly reduce the impact of I/O interference.
Figure~\ref{fig:apps} demonstrates the performance measurements of the five applications with the three settings.
With FIFO and the background job, NAMD, WRF, BERT, and SPECFEM3D are slowed down by 60.6\%, 45.3\%, 3.8\%, and 3.0\%, respectively.
With size-fair and the background job, the slowdown of each application is 0.1\%, 4.6\%, 1.6\% , and 0.0\%.
All the slowdowns of size-fair are bounded by its fair share of the I/O resource in proportion to node count.
For example, NAMD is run on 64 nodes, so the maximum possible slowdown with a background benchmark job from one node should be $1/65 = 1.5\%$, assuming NAMD is entirely I/O.
However, due to the non-trivial computation in NAMD, the measured slowdown is only 0.1\%.

The only observed exception is ResNet-50, which uses asynchronous I/O. 
The 2.7x slowdown with FIFO is reduced to 12.9\% with size-fair. 
The 12.9\% slowdown exceeds the maximum possible value of 5.9\%. which is calculated as the ratio of the background job size to the total node count of the two jobs ($1/17$).
We believe this is due to the bounding factor change: with asynchronous I/O, ResNet-50 is bounded by the computation and communication.
As the I/O latency increases, I/O becomes the dominating factor, which introduces a non-linear increase in time-to-solution.
To further validate the effectiveness of size-fair, we change the I/O of ResNet-50 to be synchronous.
Although the synchronous I/O is slower than asynchronous I/O (with a 62.1\% overhead), the size-fair policy with \name{} only introduces a 1.1\% slowdown compared to the exclusive case. 
In contrast, the FIFO policy slows ResNet-50 down by 2.0x.

The results clearly show that \name{} can dramatically reduce the impact of I/O interference by proportionally sharing I/O capability with the size-fair policy. 
With \name{} fair-share, these slowdown caused by I/O interference is reduced by 59.1--99.8\% across applications.
The resulting time-to-solutions are all below the maximum possible slowdown with the appropriate amount of I/O capability, which shows that the delay introduced by \name{} sharing policies is bounded.

\subsection{$\lambda$-delayed Fairness}
\label{sec:exp:global}
\begin{figure}[h]
\begin{center}\up\up
    \includegraphics[width=\columnwidth]{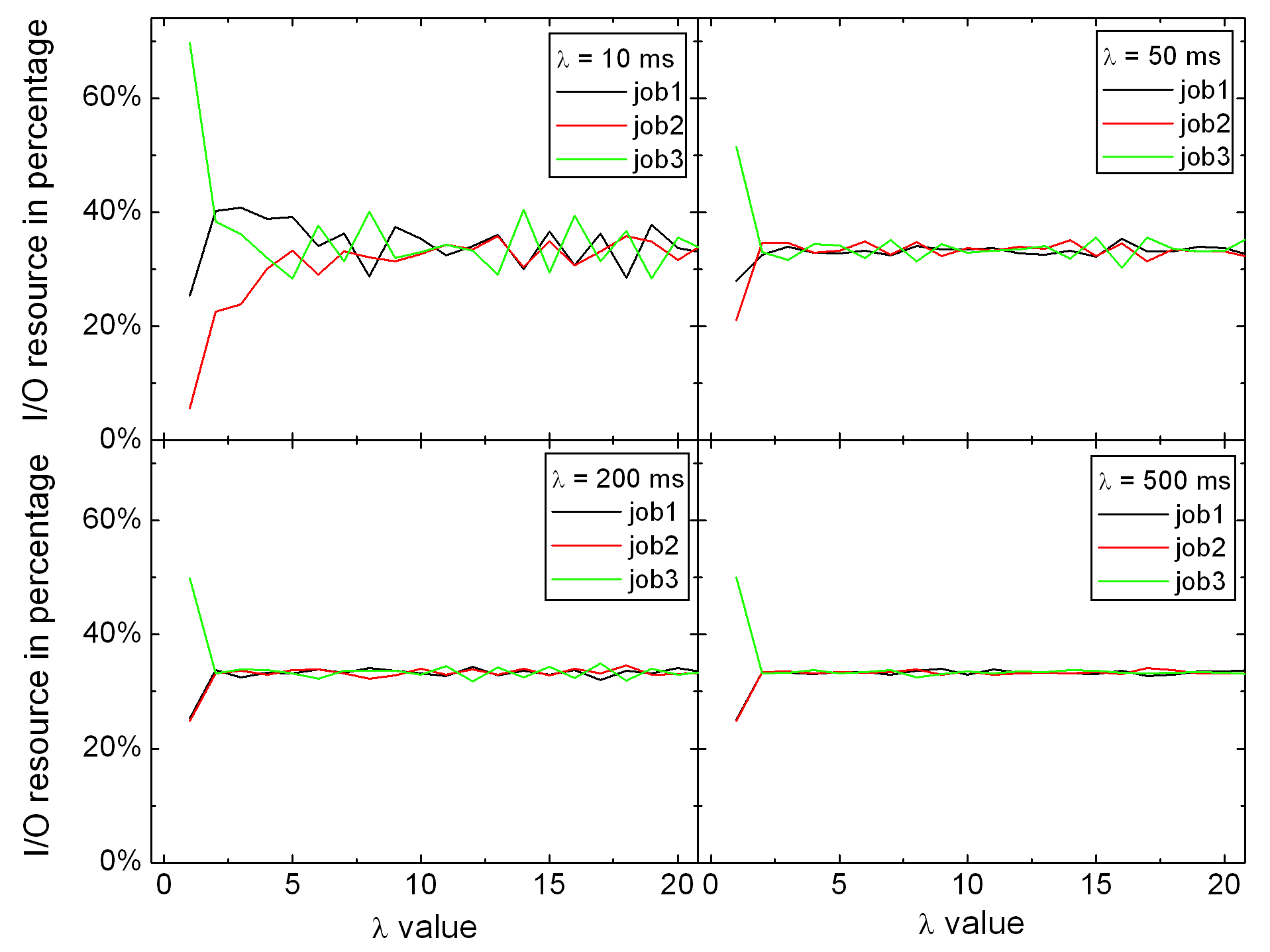}
    \up\up\up\up\caption{$\lambda$-delayed global fairness with various interval lengths ($\lambda$).}
    \label{fig:interval}\up\up\up\up
\end{center}
\end{figure}

In \S\ref{sec:design-fair}, we introduced $\lambda$-delayed fairness to mitigate an imbalanced sharing of I/O resources due to using a local job view. 
In this experiment, we vary the communication interval ($\lambda$) and study the impact of this parameter on the overall system.
This experiment has three jobs with associated files spread across two \name{} servers exclusively, so \name{} starts in a unfair sharing state.
Our tests set the communication interval to \{10, 50, 200, 500\} ms.
Figure~\ref{fig:interval} shows the sharing percentage of each job's I/O usage in the four test cases.
Using a communication interval of 50, 200, and 500 ms, \name{} reaches global fairness by the second interval. 
With a 10~ms interval, it takes five intervals for \name{} to reach global fairness.
One other thing to note is that using a shorter communication interval produces a higher variance in the I/O resources allocated among the jobs, as discussed in \S\ref{sec:design-token}. 
%This is because in each interval, \name{} assigns tokens in a probabilistic manner, and when the interval is short, the.
We observe that $\sim$50~ms is the effectiveness boundary of \name{} on Frontera.
Although this boundary depends on the processing speed of the server, the interconnect, and the underlying shared file system, we find the 500~ms communication interval is a reasonable value for real applications and benchmarks.

\section{Related Work}
\label{sec:related}
Traditional I/O research, such as MPI-IO~\cite{thakur1999implementing, thakur1997users} and ADIOS~\cite{Lofstead2008adios}, optimizes individual job performance by intelligently mapping memory region to file system data structures.
ZOID~\cite{iskra2008zoid} is an operating system level I/O component that decouples file and socket I/O and enables customized application I/O interface at scale.

Researchers have noticed interference when sharing I/O resources among applications~\cite{mubarak2017quantifying}.
In practice, production burst buffer systems, such as DataWarp~\cite{henseler2016architecture} on the Cori supercomputer, integrate with SLURM, so that users can provision burst buffers with two simple policies: {\it bandwidth} and {\it interference}. 
The {\it bandwidth policy} allocates burst buffer servers to a job to maximize its I/O throughput. 
The {\it inference policy} assigns a minimal number of burst buffer servers to a job, but with exclusive access.
%Neither policy can achieve processing isolation, Pareto efficiency, and policy proofness at the same time:
Both policies are resource underutilization prone:
The {\it bandwidth policy} can underutilize the allocated I/O resource while the {\it interference policy} can lead to resource starvation.

Other research has investigated the origin of this interference and proposed different solutions.
For example, I/O-aware job scheduling~\cite{herbein2016scalable} attributes the interference to the aggregated I/O throughput of interconnect switches, then proposes an I/O provisioning algorithm with users' input of their required throughput. 
This method guarantees that the aggregated bandwidth of jobs does not exceed that of switches; however, multiple jobs can still be placed under the same switch, which processes I/O requests in a FIFO manner and results in possible indefinite blocking of an application due to the lack of isolation.
Also, a user can easily specify a higher I/O throughput than the application needs to over-provision resources.
Physical isolation~\cite{kougkas2016leveraging} maps I/O workloads from different jobs to a disjoint group of file system servers, which provides isolation but can result in low overall efficiency as jobs may not be able to fully utilize the resources. 
In particular, this approach allocates file system servers based on the output file size specified by users, which can be gamed with boosted values.
Liang et al. analyze the impact of I/O process count on the contention problem, and propose the CARS system to map jobs to burst buffer servers to avoid such I/O contention~\cite{liang2019cars}.
Similar to I/O-aware scheduling, this approach does not support isolation, and the policy can be tampered with through user input of the I/O process count.

Recent research on I/O forwarding resource sharing such as
GIFT~\cite{patel2020gift}, TBF~\cite{qian2017configurable}, and DFRA~\cite{ji2019automatic} investigates system design and algorithms to enable efficient and fair sharing of I/O resources.
GIFT and DFRA use the fact of 80\% of HPC applications are run more than five times.
They design a specific throttle-and-reward mechanism and profile-based job placement, respectively.
In contrast, \name{} assigns I/O resources based on real-time I/O dynamics and is effective for both known and new applications.
Similar to \name{}, TBF enables I/O resource sharing among compute nodes, jobs, or I/O operations.
It requires user-supplied upper and lower I/O request rate and provides QoS accordingly.
However, it is difficult to know the exact I/O request rate of an application, even for an experienced user.
In addition, the user-supplied request rate may not be accurate.
%\name{}, on the other hand, manages I/O resources purely based on the I/O workload and provides stronger policy proofness.

\section{Conclusion and Future Work}
\label{sec:conc}
This paper has presented \name{}, an automatic, policy-driven, and efficient I/O sharing framework for burst buffer.
It enables policy-driven I/O resource sharing and minimizes the impact of I/O interference with a statistical token design to time-slice I/O request processing cycles and assigning cycles based on runtime information of jobs. 
We introduced $\lambda$-delayed fairness to mitigate the sub-optimal sharing problem due to the job information discrepancy.
We demonstrate the sharing policy flexibility of \name{} with three primitive sharing policies and two composite policies.
Our benchmark results show that \name{} can correctly and efficiently enforce specified sharing policies to assign I/O resources using various policies.
The I/O sharing enabled by \name{} sustains a 13.5--13.7\% higher I/O throughput and a 19.5--40.4\% lower performance variation than existing algorithms.
In a controlled environment, \name{} significantly reduces or eliminates the application slowdown caused by I/O interference compared to the FIFO baseline.
As future work, we are investigating various log-structure byte-addressable file system designs and persistent data structure strategy to enable fault tolerance in \name{}.

\balance
\bibliographystyle{ACM-Reference-Format}
\bibliography{ThemisIO}

%\theendnotes

\end{document}